\documentstyle[ieeemagn,epsfig]{article}
\begin{document}
\baselineskip 11.5 pt
\hoffset=.75truecm
\title{Nonlinear Response of HTSC Thin Film Microwave Resonators in an Applied DC
Magnetic Field}
\author{Durga P. Choudhury${}^{a,b}$, Balam A. Willemsen${}^{a,b}$, John S.
Derov${}^b$ and S. Sridhar${}^a$\\{\small ${}^a$Physics Department,
Northeastern University, Boston, MA 02115}\\{\small ${}^b$Rome Laboratory,
Hanscom AFB., Bedford, MA 01730}}
\thanks{Manuscript received August 27, 1996.
\newline \indent D.\ P.\ Choudhury, 617-373-2948, fax 617-373-2943,
dpc@neu.edu, http://sagar.physics.neu.edu/
\newline \indent Work at Northeastern University was supported by the AFOSR
through Rome Labs, Hanscom AFB.}
\maketitle

\abstract{
The non-linear microwave surface impedance, $Z_s=R_s+iX_s$, of patterned
YBCO thin films, was measured using a suspended line resonator in the
presence of a perpendicular DC magnetic field, $H_{DC}$, of magnitude
comparable to that of the microwave field, $H_{rf}$. Signature of the virgin  state
was found to be absent even for relatively low microwave power levels. The
microwave loss was initially found to {\em decrease}\/ for small applied
$H_{DC}$ before increasing again. Also, non-linearities inherent in the
sample were found to be substantially {\em suppressed} at low powers at these applied
fields. These two features together can lead to significant improvement in
device performance.
}

\section{Introduction}

The microwave response of high-T${}_c$ superconductors (HTSC) is important
both from the point of view of microwave applications of HTSC\cite{ZYShen94a}
and fundamental physics\cite{TCLGSollner96a}. An understanding of the loss
mechanisms, field and current profiles and nature of non-linearities can
lead to improvement in fabricated devices that use them. While numerous
experimental studies of non-linear microwave response of HTSC have been done
\cite{CWilker95a}, none of them to our knowledge have probed the
non-linear response in the presence of DC magnetic fields, where the DC and
microwave fields are of comparable magnitude. In this situation the effect
of the microwave field cannot be considered as a small perturbative Lorentz
force on the vortex lattice generated by the DC field, as is often done at
high fields\cite{MWCoffey92a}. Such a situation can also act to test
various models that have been proposed for the non-linear response.

\section{Experimental Techniques}

We used a patterned suspended resonant thin film of YBa${}_2$Cu${}_3$O$
{}_{7-\delta }$ housed in a rectangular copper package to carry out this
series of experiments. Similar methods have been used before\cite
{BAWillemsen94b,BAWillemsen95a} with great success to investigate vortex
dynamics and non-linearities in HTSC. The film, procured from Neocera Inc,
was deposited on a 0.6~in $\times$ 0.22~in $\times$ 0.010~in LaAlO${}_3$ substrate by
laser ablation, and was subsequently patterned in-house to a
straight line of dimension 0.56~in $\times$ 0.004~in using methods described elsewhere
\cite{BAWillemsen95t}. In order to obtain the highest possible $Q^{\prime }$%
s, the package was mechanically polished and thoroughly cleaned before
each set of experiments. The resonator was made symmetric by placing a blank
substrate of the same material and dimensions on top of the film before it
was loaded into the package.

The assembly, complete with a controlling heater and temperature sensor, was
inserted into the sample chamber of a Cryo Industries Variable Temperature
cryostat. Two independent
carbon glass sensors and temperature controllers were used to stabilize the
temperature of the cavity to the degree required for these experiments. A
LakeShore DR91C was used for gross control of the vaporizer temperature,
which was set slightly below the desired sample temperature. The desired
sample temperature was obtained and finely controlled with a LakeShore
DR93CA. Temperature stability of the order of 1~mK were typical for the
experiments presented here, where the data took up to two hours to obtain
for each run.

DC magnetic fields up to $~$1000~Oe was applied parallel to the $c$-axis of the
sample using a custom-built Walker Scientific copper solenoid and a
LakeShore 622 superconducting power supply. Unlike typical Helmholtz coil
configurations, the solenoid has no polecaps, thus ensuring that there is no
remanent field, save for possibly the geomagnetic fields. It is worth
pointing out that our experiment does not use any superconducting ground
planes unlike parallel plate or microstrip resonators, thus avoiding
complications due to demagnetization effects from such plates.

Microwaves were inductively coupled to and from the resonator by means of
loops at the ends of stainless steel coaxial lines. The microwave
transmission amplitude $S_{21}$ was then measured using an HP 8510C
Automatic Network Analyzer. The coupling strength was adjusted by varying
the distance between the loops and the resonator as well as their relative
orientation. Coupling could thus be reduced to the point that the loaded
and unloaded $Q^{\prime }$s are indistinguishable, simplifying the data
extraction process. At the low input power levels that were used to carry
out these experiments, the trace was very noisy. This coupled with the fact
that we require extreme sensitivity to very small changes mean that we could
not simply determine the $Q$ from the maximum frequency and $-3$dB bandwidth
as is often done. In order to reduce the noise and obtain the required
sensitivity:

\begin{itemize}
\item  The network analyzer was used in the ``Step'' mode, in which every
frequency point is individually synthesized

\item  The signal was heavily averaged to get rid of the random noise,

\item  The trace was fitted using the method of least squares to Lorentzian
shape, and

\item  The frequency span was kept as narrow as possible, usually only about
20\% larger than the $-3$~dB bandwidth.
\end{itemize}

The center frequency and the $-3$dB bandwidth obtained from the fit agreed
very well with those directly read off the trace, especially at low power
levels where the trace is closest to Lorentzian shape, but provided
significantly enhanced sensitivity to small changes.

{
\begin{center}
\begin{figure}
\mbox{\epsfig{file=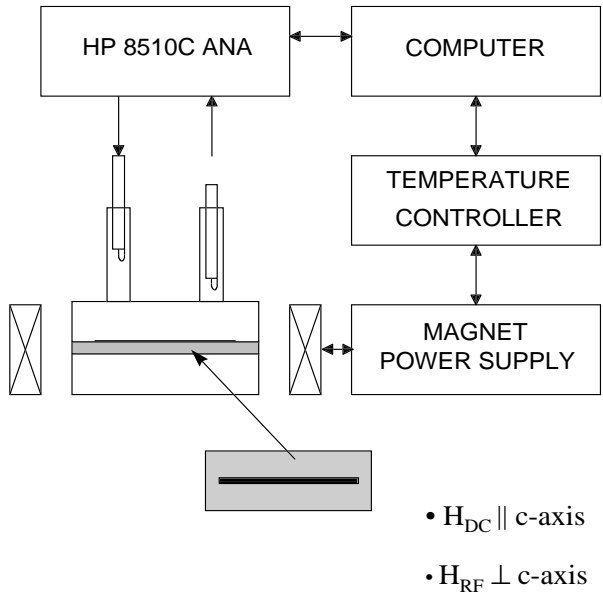,height=3.2truein,width=3.2truein,angle=0}}
\caption{Block diagram of Experimental Setup}
\label{fig1}
\end{figure}
\end{center}
}

\section{Results and Discussion}

Ubiquitous intrinsic non-linearities have been observed in thin film
specimens of High-T${}_c$ materials\cite{PPNguyen93a,BAWillemsen95a}, and
some aspects of these non-linearities appear to be explained by a
current-induced critical state model\cite{SSridhar94a}. The present
experiments, which involve both microwave and DC fields of comparable magnitude
so as to study the interplay of these two effects, were designed to further
test these critical state and other ideas. Our experiments show that
the presence of even relatively low microwave powers can induce vortices in
the film, emulating the response of a DC field. The signature of this fact
come from the observation that low DC field hysteresis does not show the
virgin state response.

In a typical sample, the signature of the virgin state (i.e. absence of trapped flux tubes) in
the low DC field hysteresis experiments manifests itself as a sharp rise in
the $-3$dB bandwidth as field is slowly increased from zero corresponding to
initial penetration of flux. As the field is further increased to a value $H_{max}$ and then cycled between $H_{max}$ and $-H_{max}$, where $H_{max}$
is a field of the order of a few hundred Gauss, this initial behavior is never reproduced; instead, it goes through a butterfly-shaped hysteresis loop\cite{BAWillemsen96a}.

The same experiment, performed on the films under discussion, yields two new
observations :

\begin{itemize}

\item The initial ``virgin'' response vanishes at higher microwave powers.
This seems to indicate that microwave fields can create enough vortices in
in the sample to wash away the virgin state response, mimicking
the effect of an applied DC field.

\item A sharp dip in $R_s$ is observed at a field scale $H_{DC}\sim 5$G in the
virgin response, indicating that a small applied DC field serves to {\it lower\/}
$R_s$.

\end{itemize}

{
\begin{center}
\begin{figure}
\mbox{\epsfig{file=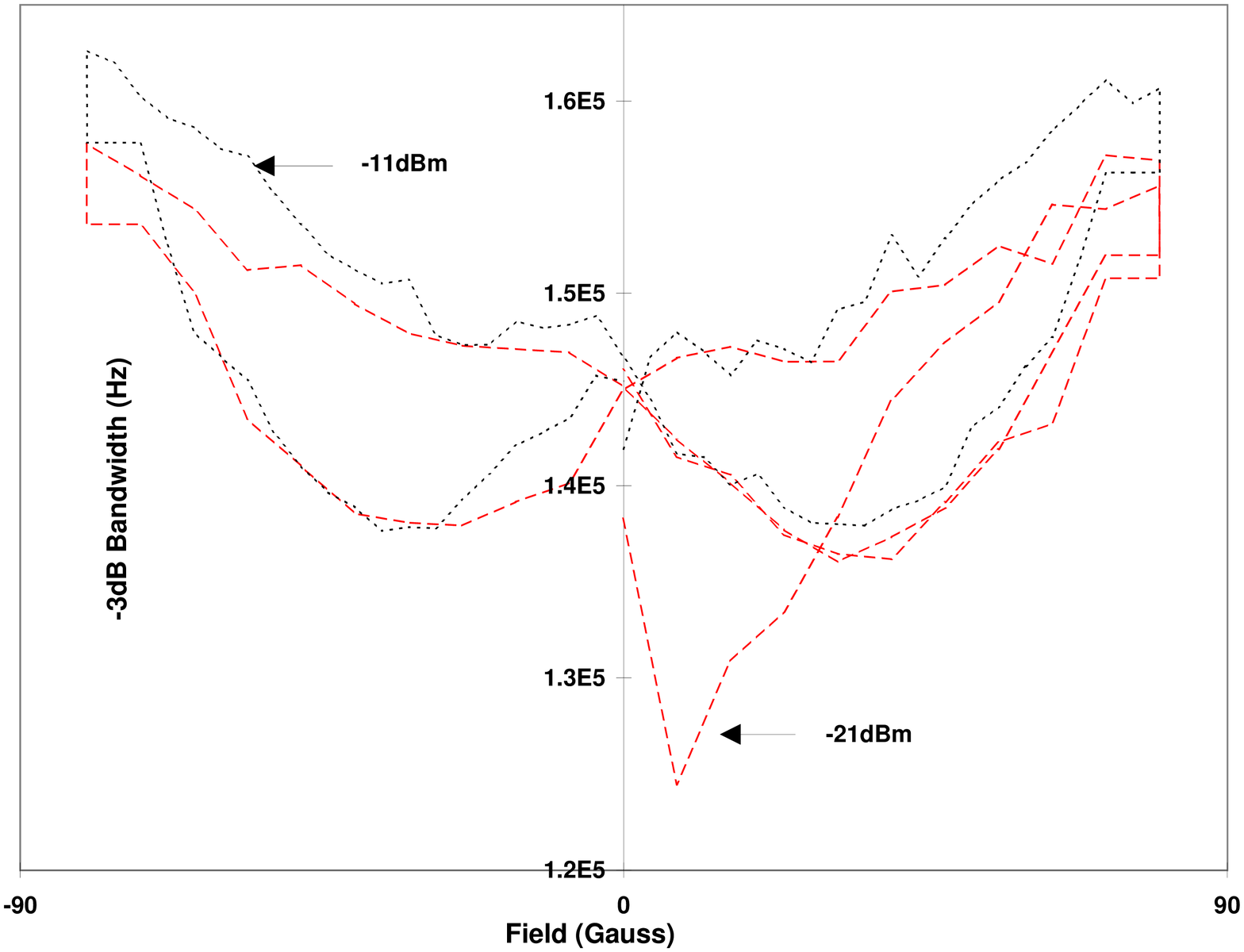,height=3.2truein,width=3.2truein,angle=0}}
\caption{Low field hysteresis at $-21$ dBm and $-11$ dBm of input power,
	 at 10K. Notice the absence of the ``tail'' corresponding to virgin
	 response at higher powers. In order to highlight the similarities
	 between the two plots, they have been superposed on each other by
	 adding a constant of 10 kHz off the $-21$ dBm power plot.}
\label{fig2}
\end{figure}
\end{center}
}

The second result was verified when we did a measurement of $R_s$ against applied
microwave power in a fixed $H_{DC}$. The decrease in $R_s$ reproduces itself,
as is evident from fig.3. Another observation from the microwave
power ramp experiments is that the non-linearities in the sample also get
substantially suppressed at these low field for low microwave powers. As
$H_{DC}$ is increased, $R_s$ gradually rises and finally goes above it's
zero field value.
To further ensure that this
observation is genuine and is not an artifact of some experimental
inconsistency, we repeated the measurements on a film patterned out of a
different albeit similar wafer. Although results obtained from this film  were not quantitatively identical with those of the other, which would be expected because
of the differences on growth, deposition and patterning of the two films, the
two characteristic features described above was observed to a comparable
extent in the second film. Also to rule out the effect of any stray
remanent DC field, we carried out the microwave power ramp measurements
with the DC field reversed and no such effects were found.

{
\begin{center}
\begin{figure}
\mbox{\epsfig{file=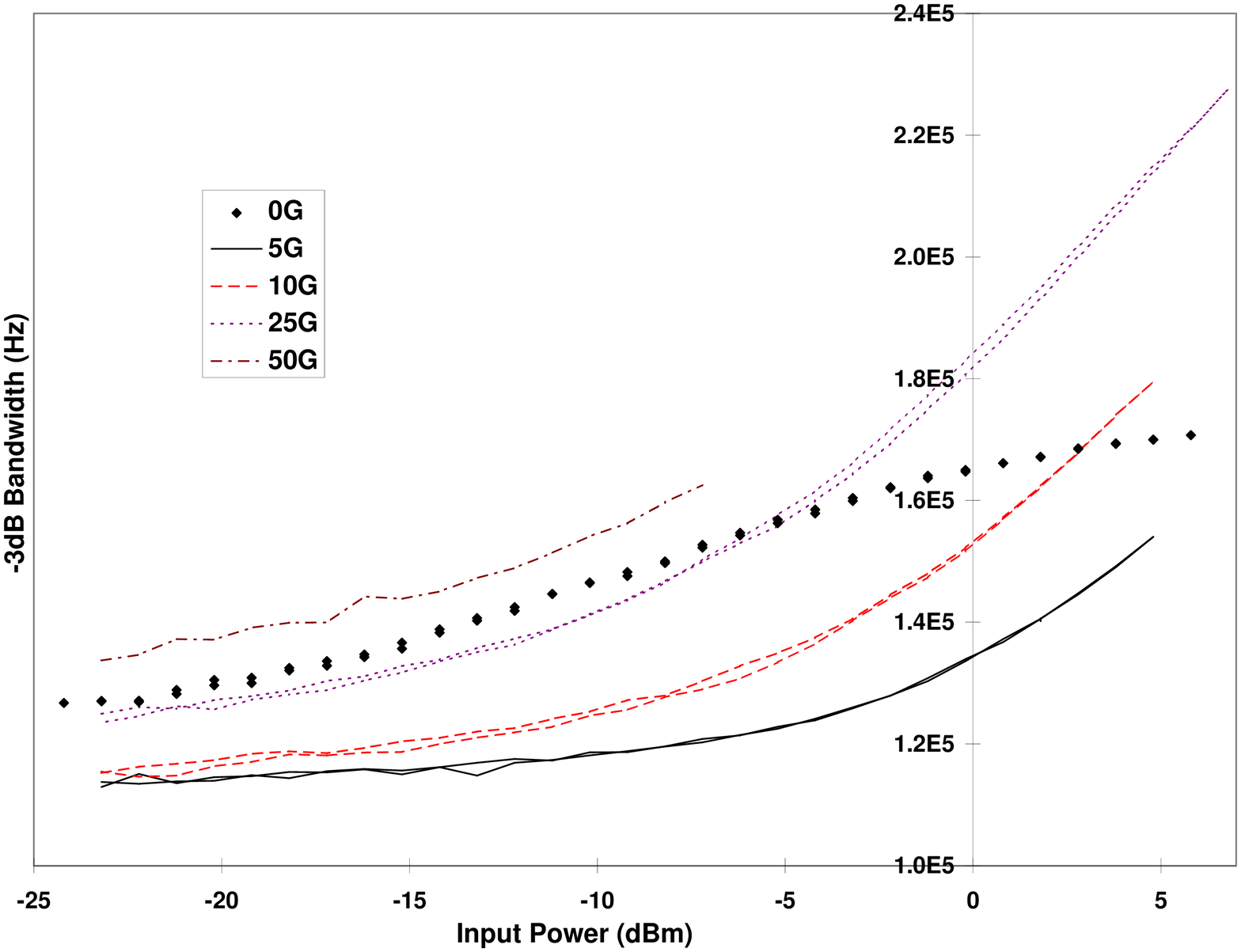,height=3.2truein,width=3.2truein,angle=0}}
\caption{Typical power dependence of resonance widths, taken at 10K}
\label{fig3}
\end{figure}
\end{center}
}

We have examined the present data in the framework of two proposed
explanations for non-linear response in HTSC, viz. weak links and dynamics
of a current-driven critical state. The weak link picture can be viewed in
terms of a resistively shunted junction (RSJ) model, taking the effect of
the DC field to be a DC current flowing on the surface in addition to the rf
current. The equation of motion for the relative phase between the
coupled grain then become $\dot{\phi}+\sin \phi =i_{dc}+i_\omega \cos \omega
t$. The dynamic impedance can be calculated from $Z_\omega =\dot{\phi}
_\omega /i_\omega $. While numerical calculations of this response yield
very interesting effects as $i_{dc}$ is varied, this approach does not seem
to yield the results that are observed in this experiment.

The critical state model should also lead to ac + dc effects, so that the
non-linearity should be modified by the DC magnetic field. However a
calculation of this effect is not straight-forward, since it requires a
prescription for the present case where the loss needs to be calculated when 
$i_{rf}$ is varied over on rf cycle for finite $H_{DC}$. The available
prescription  in the literature \cite{EHBrandt93a} does not consider this
case, but instead considers a different method of varying $i_{rf}$ and $H_{DC}$.
Hence, at the present, it is not possible to determine if the critical state
model can explain these unusual results.

It is worth noting that the unusual decrease in $R_s$ observed here can occur due
to non-equilibrium effects and in fact have been seen in low $T_c$
superconductors \cite{SSridhar83t}. There it was shown that when the
microwave frequency $\omega >\tau ^{-1}$, where $\tau $ is the quasiparticle
relaxation time, a non-equilibrium quasiparticle distribution can occur
which leads to a decrease of $R_s$ in the presence of an $i_{dc}$. Another
related phenomenon which occurs is an enhancement of the superconducting
gap. While this condition is met in pure metals at low temperatures, it is
not clear if this happens in the high $T_c$ materials. 

However it is interesting to note that in YBa${}_2$Cu${}_3$O${}_{7-\delta }$ crystals, 
$R_s(T)$ is non-monotonic and there are regions of temperature where $%
(\partial R_s/\partial T)<0$. This unusual, apparently non-thermodynamic
result, may imply that $(\partial R_s/\partial i_{dc})<0$ need not be
surprising.

\section{Conclusion}

We have described a novel effect in which both the microwave losses and
non-linear response decrease in the presence of small magnetic fields.
Although a clear explanation of this effect is lacking, and it could arise
from non-equilibrium quasiparticle effects, the present observation implies
that losses can be reduced by as much as $30\%$ and could have interesting
implications for device performance.

\bibliographystyle{ieee}
\bibliography{strings,big}
\end{document}